\documentclass[12pt,aps,prb,preprint]{revtex4}   

\usepackage{amsmath}    
\usepackage{graphicx}   

\begin{document}

\title{Investigating viscous damping using a webcam}
\author{Sohaib Shamim}
\email{sohaibshamim@lums.edu.pk}
\author{Wasif Zia}%
 \email{wasif.zia@lums.edu.pk}
\author{Muhammad Sabieh Anwar}%
 \email{sabieh@lums.edu.pk}
\affiliation{School of Science \& Engineering, Lahore University of
Management \& Sciences (LUMS), Opposite Sector U, D.H.A, Lahore
54792, Pakistan}\date{\today}

\begin{abstract}
We describe an experiment involving a mass oscillating in a viscous fluid and analyze viscous damping of harmonic motion. The mechanical oscillator is tracked using a simple webcam and an image processing algorithm records the position of the geometrical center as a function of time. Interesting information can be extracted from the displacement-time graphs, in particular for the underdamped case. For example, we use these oscillations to determine the viscosity of the fluid. Our mean value of $1.08$ $\pm$$0.07$ mPa\,s for distilled water is in good agreement with
the accepted value at $20^\circ$C. This experiment has been successfully employed in
the freshman lab setting.
\end{abstract}

\maketitle
\section{\label{sec:intro}INTRODUCTION}

The topics of oscillatory and simple harmonic motion (SHM) are of fundamental importance to physics and engineering. The differential equations that lie at the heart of oscillatory motion are frequently covered in introductory undergraduate courses. The applications are also widespread, touching upon phenomena as diverse as vibrations of atoms in a crystal, current in LCR circuits, the transmission of electromagnetic radiation in dielectrics and chaos. However, we felt that even though a large number of our students knew how to solve these second order differential equations, very few had actually seen the solutions naturally emerge in real physical situations, especially in the formal laboratory environment. The present experiment demonstrates these solutions and provides physical insight into light and heavy damping. The data, that is acquired from an off-the-shelf webcam, is subsequently processed to calculate a useful parameter, the viscosity of commonly available fluids such as distilled water, ethanol and methanol. The values are in reasonable agreement with accepted values.

In the process, students are also exposed to techniques in image processing. The use of these techniques is becoming more widespread especially in the area of video microscopy where a CCD camera is used to track the motion of particles, microspheres or fluorescent proteins in fluidic environments. Some of these ideas have also been discussed\cite{nakroshis,jia} in this journal and used to follow Brownian motion and its dependence, for example, on fluid viscosity, particle size and temperature as well as the estimation of Boltzmann's and Avogadro's constants\cite{newburgh}. The present experiment may serve as a first introduction to some commonly used algorithms and tools in image processing such as frame grabbing, color control, motion tracking and using videos for the quantitative verification of mathematical and semi-empirical models.

Not surprisingly, there is already a diverse r{\`{e}}pertoire of experiments performed in instructional physics laboratories worldwide that analyze different facets of harmonic motion. These include the use of oscillating water columns \cite{gilson}, swinging pendulums \cite{mcinerney} and masses attached to springs. These experiments analyze SHM as well as its nonlinear generalizations using potentiometers \cite{mcinerney}, photocells \cite{allen}, photosensors \cite{gupta} and force sensors \cite{alexander}. In the present experiment, we use a webcam to track the damping of a mass oscillating in various fluids such as honey, water, ethanol and methanol and adjust the fluids to exhibit heavy and light damping.

Several illuminating articles revolving around the concept of viscous damping have appeared from time to time in pedagogical journals. In our case, we process the images acquired from the webcam, in the case of light damping, and go on to calculate the \textit{absolute} viscosity of the fluid, a parameter that is conventionally determined using viscometers of various kinds \cite{chung}. Our proposed technique is a new method of acquiring data from a viscously damped system and differs from traditional methods in several useful respects. For example, Alexander and Indelicato \cite{alexander} used a force sensor to monitor the damping of a spherical mass in water and processed their data to determine the viscosity. Their calculation differed from the accepted value by an order of magnitude, indicating the presence of large, systematic errors. Another article\cite{gupta} described the use of a photosensor to track the viscous damping of a pendulum in air. This method works very well for gases but the downside is that ordinary photosensors cannot be immersed in liquids. Our method employing the webcam acquires data \textit{remotely} obviating the need for any physical contact with the oscillator. The positions are directly measured and the time stamped displacements can also be numerically processed, if desired, to estimate the velocities\cite{pecori}. We show that in conjunction with the appropriate mathematical model, ours is a quantitatively accurate method for determining the viscosity of liquids.

The present experiment extends the list of webcam-based experiments oriented for undergraduate labs demonstrating the diffraction of light, the  diffusion of ink in water\cite{nedev}, as well as quantitative measurements on shadows, sprouting water jets, hanging chains and caustic reflections \cite{gil}.

\section{\label{sec:theory}THEORY}
The physical model is of a linear harmonic oscillator,
a spherical mass attached to a spring that remains within its
elastic limit and executes simple harmonic motion inside a viscous
medium that we call the fluid. At small speeds, a good approximation
to the damping force is given by\cite{french},
\begin{equation}
\label{eq:opposingforce} F \approx bv,
\end{equation}
where $v=\dot{y}$ is the linear speed of the oscillator and $b$ is a
proportionality constant characterizing the medium and the shape of the
oscillator. The characteristic
equation of motion is,
\begin{equation}
\label{eq:dampedshm}
\ddot{y} + \gamma\dot{y} + \omega^2y = \xi,
\end{equation}
where $\gamma$ = $b/m$, $\omega^2$ = $k/m$, $k$ is the spring
constant, $m$ is the mass of the particle and $\xi$ is the ratio of
the buoyant force to the mass of the sphere. The constants $\gamma$
and $\omega$ represent, respectively, the damping constants and the
natural frequency of the oscillator when the damping is switched
off. During the experiment, the oscillator is completely
immersed in the fluid at all times. As a result, the buoyant force is constant and always directed upwards.
Therefore, $\xi$ is a constant and causes a uniform offset in the
displacement $y$; mathematically, this is a shifting of the
solutions of Equation \eqref{eq:dampedshm} on the vertical axis (in a displacement-time graph).

A general expression for the damping force experienced by a
spherical mass of radius $r$, oscillating in a fluid of viscosity
$\eta$ and density $\rho$ is given by
\cite{lifshitz,leshansky,gupta},

\begin{equation}
\label{eq:landau} F = - 6\pi\eta r\biggl(1 +
\dfrac{r}{\delta}\biggr)v - 3\pi r^2\biggl(1 +
\dfrac{2}{9}\dfrac{r}{\delta}\biggr)\rho\delta \dfrac{dv}{dt},
\end{equation}
where $\delta$ is called the penetration depth,
\begin{equation}
\label{eq:deltadef} \delta = \sqrt{\dfrac{2\eta}{\rho\omega}}.
\end{equation}
The penetration depth defines the depth at which the wave amplitude
falls to $1/e$ (about $37\%$) of the amplitude at the surface of
the sphere. This depth decreases with the frequency of the wave, but
increases with the kinematic viscosity $\eta/\rho$ of the fluid.
From Equation \eqref{eq:landau}, we recover Stoke's famous law, $F = -
6\pi\eta rv$, if the mass is falling with uniform velocity and
$\omega$ is zero. If the dimensions of the sphere are greater than the skin depth, $r$$\gg$$\delta$, Equation~\eqref{eq:landau} reduces to,

\begin{equation}
\label{eq:landauapprox} F \approx - \dfrac{6\pi\eta r^2}{\delta}v -
\dfrac{2\pi r^3}{3}\rho \dfrac{dv}{dt},
\end{equation}
but the expression applies if the mass is oscillating in a container of infinite dimensions; for the physically realizable case of a finite container,
a correction term is added \cite{shantarng},

\begin{equation}
\label{eq:landauapproxfinite} F \approx - \biggl(\dfrac{6\pi\eta
r^2}{\delta}v + \dfrac{2\pi r^3}{3}\rho
\dfrac{dv}{dt}\biggr)\biggl(1+2.1\dfrac{r}{d}\biggr),
\end{equation}
$d$ being the radius of the cylindrical container. Now from
\eqref{eq:opposingforce} and \eqref{eq:landauapproxfinite}, it can
be shown that the oscillating mass obeys the equation of motion
\eqref{eq:dampedshm}
with the mass $m$ replaced by $m_\textrm{f}$ and the parameterizations,
 \begin{equation}
 \label{eq:gamma}
 \gamma = \dfrac{6\pi\eta r^2}{m_\textrm{f}\delta}\biggl(1+2.1\dfrac{r}{d}\biggr),
 \end{equation}

 \begin{equation}
\label{eq:mf}
 m_\textrm{f} = m_{\textrm{sphere}}+\dfrac{2\pi r^3}{3}\rho\biggl(1+2.1\dfrac{r}{d}\biggr) \quad\quad\text{and}
\end{equation}

 \begin{equation}
 \label{eq:omega}
   \omega^2  = \dfrac{k}{m_\textrm{f}} .
 \end{equation}
From Equation ~\eqref{eq:mf} we deduce an interesting fact: if the container is infinitely large, exactly half of the
fluid displaced by the sphere accompanies it in its oscillatory
journey by forming a boundary film around the sphere! We estimate the damping coefficient from the mass that negotiates underdamped SHP and use the relationship~\eqref{eq:gamma} to estimate the viscosity. Finally, as a simple demonstration, students employ solutions, such as honey in water, adjust their concentrations and investigate light and heavy damping.

\section{\label{sec:experiment}THE EXPERIMENT}
\begin{figure}
\includegraphics [width=9.0cm]{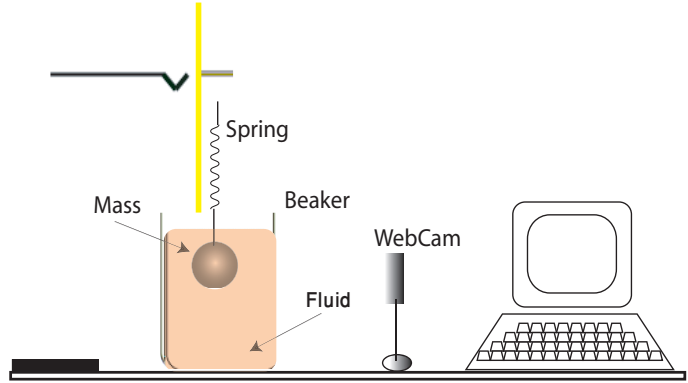}
\caption{Schematic diagram of the experimental
set-up. The apparatus is adjusted so that the mean position of the spherical mass is in the horizontal line of sight of the webcam. \label{schematic}}
\end{figure}

\begin{figure}[!h]
\includegraphics [scale=0.50]{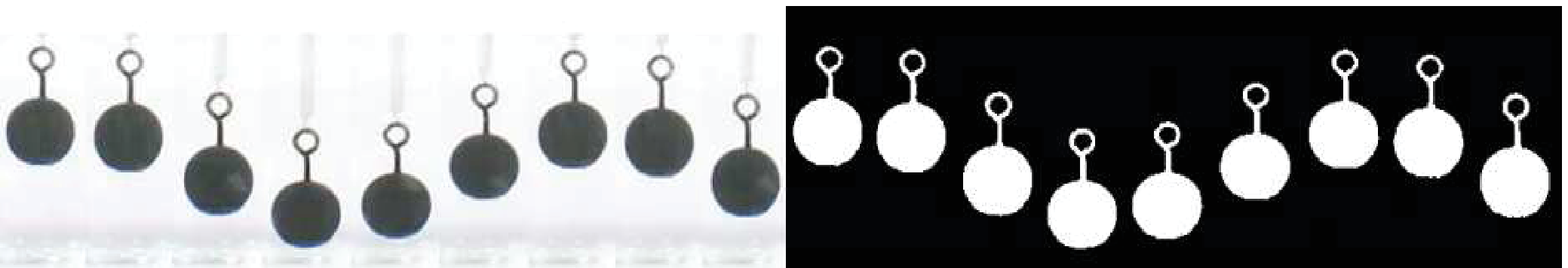}
\caption{(a) Colored photographs of the frames showing the
oscillations. (b) Typical black and white frames after
processing. The images are cropped to show only the region of interest.\label{frames}}
\end{figure}

The experimental setup is simple, easily repeatable and is illustrated in
Figure~\ref{schematic}. A sphere of mass $210~$g is attached to
a spring of spring constant $8~$Nm$^{-1}$ and set to oscillate vertically in
the fluid while a webcam acquires a time-series of frames.
In our case, the \textit{nominal} frame acquisition rate is set at $30$ per
second with $300$ frames acquired in total. A white sheet of cardboard is placed in the background to
help reduce noise levels while the spherical mass is also painted black
also that helps in further reducing the light reflections. Figure~\ref{frames}(a) shows a string of acquired colored frames. To reduce the noise levels even further, the frames are cropped showing only the interesting region in which the mass oscillates.
The image and data processing are all done in Matlab. However, this
is only out of convenience and the students' prior training---any
other commercially available software can be used with equal
ease. The data analysis is described in the next section.

After the frames have been acquired, the next step is to calculate the centroid of the oscillating sphere. This a common technique in image processing and Matlab provides an algorithm for the purpose \cite{matlab}. Since
the accurate determination of the centroid is susceptible to color fluctuations,
the frames are converted to black and white at this stage with the
mass appearing white against a black background. Figure~\ref{frames}(b)
shows a string of black and white, cropped frames, corresponding to the colored ones in Figure~\ref{frames}(a). The centroids of
the white regions in Figure~\ref{frames}(b) are determined for all
the $300$ frames and then plotted as a function of time as we now
discuss.

\section{\label{sec:data}DATA ANALYSIS}
\begin{figure}[h]
\includegraphics [scale=0.5]{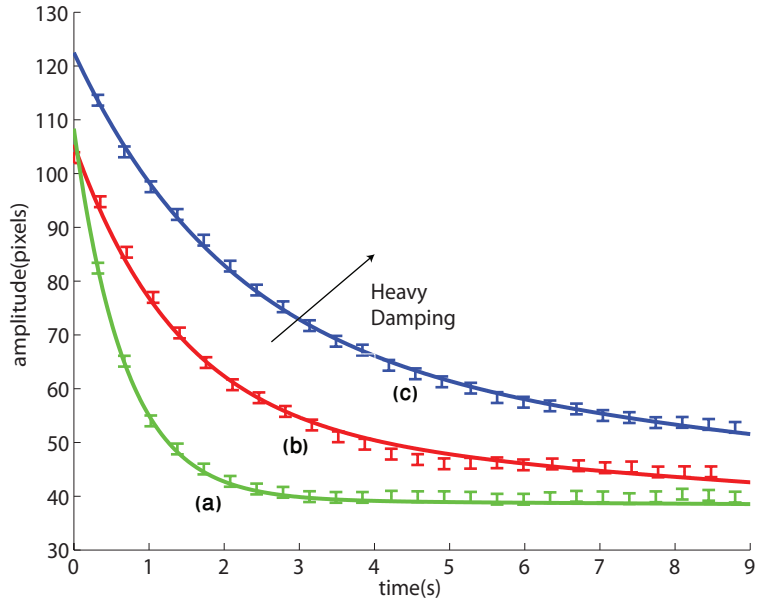}
\caption{Damping in a jar of honey depicting overdamped SHM and the varying rates of approach towards equilibrium. The
parameters are (a) $\mu_1=1.475~$s$^{-1}$, $\mu_2=0.002~$s$^{-1}$, $A=69~$p and $B=39~$p; (b) $\mu_1=0.711~$s$^{-1}$, $\mu_2=0.021~$s$^{-1}$, $A=54~$p and $B=51~$p; and (c) $\mu_1=0.484~$s$^{-1}$, $\mu_2=0.025~$s$^{-1}$, $A=59~$p and $B=64~$p.\label{heavydamps2}}
\end{figure}

The solutions of the differential equation \eqref{eq:dampedshm} reveal three different kinds of harmonic motion depending on the relative magnitudes of
$\gamma$ and $\omega$. For example, the condition $\omega^2$ $<$ $\gamma^2/4$ represents overdamping (or heavy damping). We experimentally achieve this regime  by oscillating the mass in a jar of honey whose density is empirically adjusted between
$1.28$ and $1.35~$g cm$^{-3}$. This case is exhibited in Figure~\ref{heavydamps2}, showing the displacement $y$ as a function
of time, the curves (a) through (c) represent increased concentrations and densities of the honey mixture, resulting in heavier damping. The viscous drag of the thicker fluid causes the mass to only slowly approach equilibrium, without performing any
oscillations. As the plots show, the approach towards equilibrium becomes slower from (a) to (c), i.e., with increasing damping coefficient. The general form of the heavily damped solution is given by\cite{main},

\begin{equation}\label{eq:solution-hd}
y(t) = Ae^{-\mu_1t} + Be^{-\mu_2t},
\end{equation}
where
\begin{equation}
\mu_{1,2} = \dfrac{1}{2}\gamma \mp
\sqrt{\dfrac{1}{4}\gamma^2 - \omega^2}.
\end{equation}

 \noindent The constants $A$ and $B$ have dimensions of length and depend on the initial state (position and velocity) of the mass, where $t=0$ is defined by the registration of the first frame. In our case, $A$ and $B$ are determined from fitting the displacement-time curve to the solution \eqref{eq:solution-hd}. 

\begin{figure}
\includegraphics [scale=0.1]{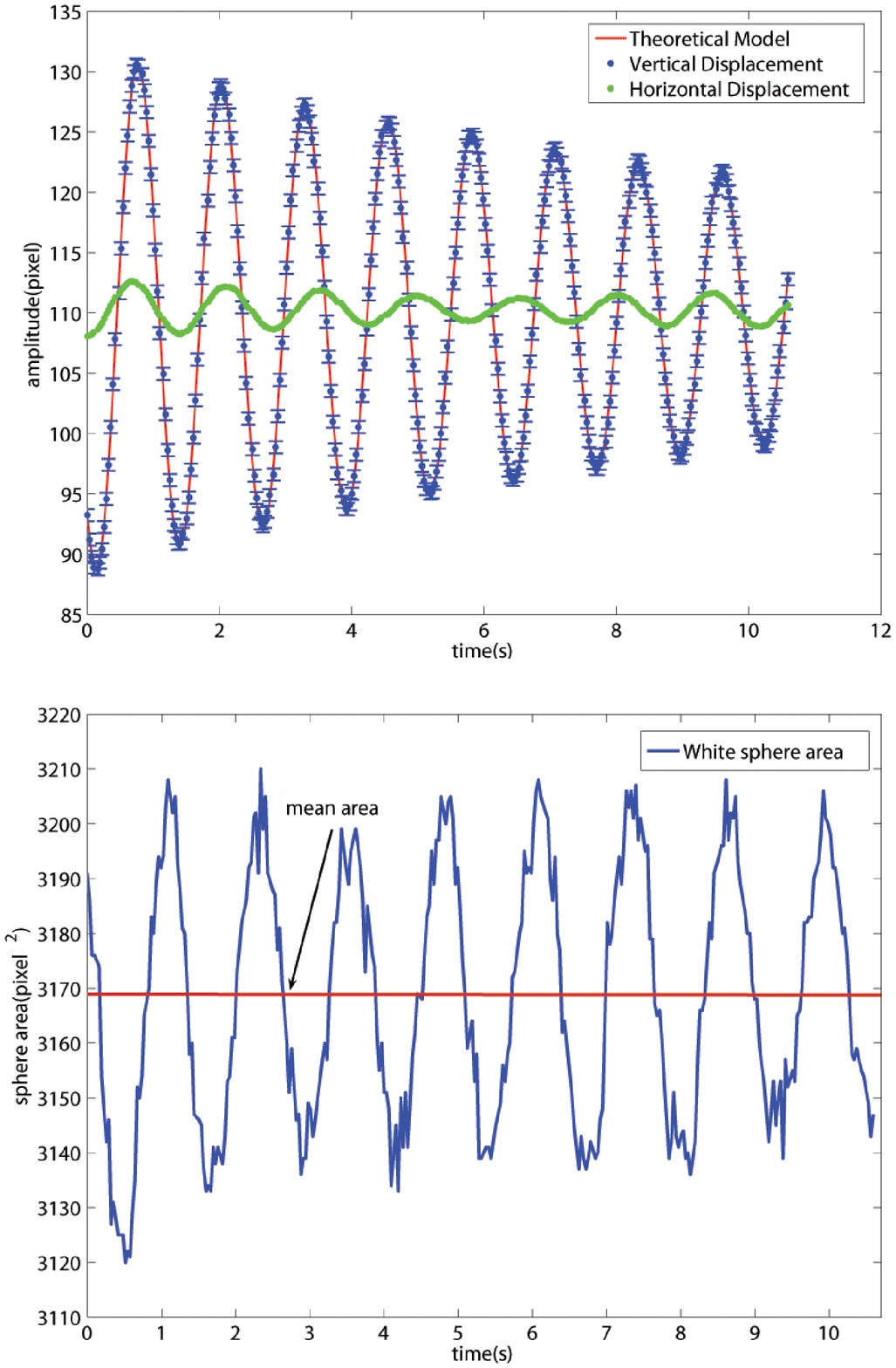}
\caption{(a) Damped simple harmonic oscillations versus time. The fit parameters are $A=21.4~$pixels, $\phi=3.82~$radians and the decay constant, $\gamma$=0.1398 $s^{-1}$. (b) Fluctuations in area of the white sphere versus time.\label{fig:underdamped}}
\end{figure}

%

The case of light or underdamped SHM, $\gamma$ $\ll$ $\omega$ is achieved when the mass is made to oscillate in a less viscous medium such as
distilled water or pure ethanol. The angular frequency of the mass $\omega_w$ is then given
by \cite{main}, 
\begin{equation}
\omega_w = \sqrt{\omega^2 - \dfrac{\gamma^2}{4}} = \omega\sqrt{1 -
\biggl(\dfrac{\gamma}{2\omega}\biggr)^2}
\end{equation}
and the corresponding displacement becomes,
\begin{equation}\label{eq:xdamp}
 y(t) = Ae^{-\gamma t/2}\cos(\omega_wt + \phi),
\end{equation}
where $A$ and $\phi$ are, respectively, the amplitude and phase determined from the initial conditions.

Figure~\ref{fig:underdamped}(a) shows the experimental data points along with
the predicted theoretical model, encapsulated by the expression \eqref{eq:xdamp}. The data points are fit to the solution and the parameter $\gamma$, representing the decay of the oscillations, is determined. Finally, the viscosity is calculated from Equation \eqref{eq:gamma}. For distilled water, our value of
$\eta$ equals ($1.08\pm0.07$) mPa\,s and is in good agreement with
the accepted value of $1.002$ mPa\,s at $20^\circ$C. For ethanol and methanol, the determined viscosities are $(1.20\pm 0.07)~$mPa\,s and $(0.39\pm 0.07)~$mPa\,s, concordant with the viscosities determined by other methods\cite{landolt,xiang}, $1.1005~$mPa\,s and $0.53~$mPa\,s.

\begin{figure}[!h]
\includegraphics [scale=0.4]{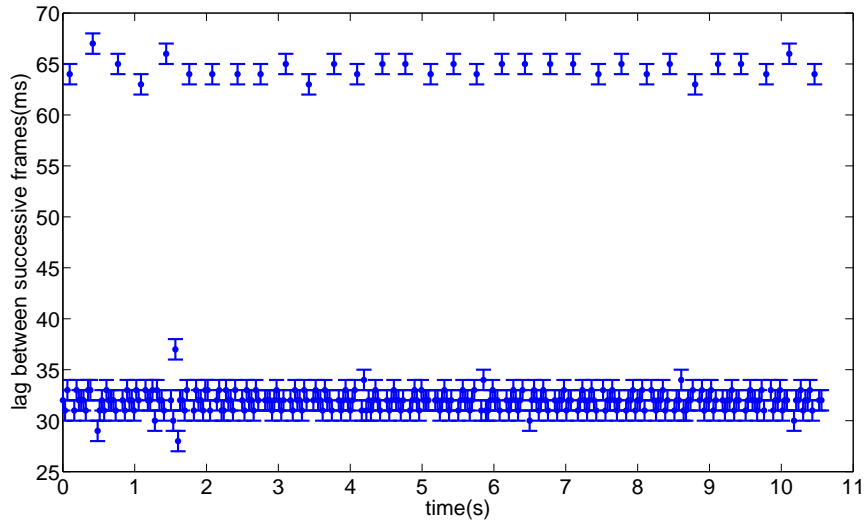}
\caption{Plot of the time lapse between successive frames in a typical run of the experiment. It is clear that the recorded frames cluster in two groups. One is the group of frames that are acquired at or near the nominal frame rate of $1/30\approx 0.033~$s whereas the second is the group of frames that are delayed by twice the sampling time, $\approx 0.066~$s; the latter represent dropped frames. We use the actual time values in our experiment instead of an average time.\label{fig:frametimediff}}
\end{figure}

The uncertainty in the results can principally arise from two sources: uncertainties in time and in position. With a digital camera, frames are dropped or delayed and some frames can even arrive earlier than expected. These variabilities give rise to timing jitter. We circumvent this problem by using the real time stamps instead of the nominal inter-frame delay of $1/30~$s (inverse of the frame rate)\footnote{This is readily achievable through the inbuilt Matlab functions (in our case, we use \textbf{getdata}).}. In case a frame is dropped or delayed, the time data explicitly shows that. The time stamps for a typical run of the experiment are shown in Figure~\ref{fig:frametimediff} and reveal the inter-frame variation. Finally, the least count of the time measurements as resolved by our software is $\pm$1 millisecond.

The uncertainty in position can be caused by multiple sources. One source is the spurious reflections of light from the metallic sphere, the glass of the beaker, refraction and the lighting conditions inside the room. For example, while computing the centroids in the acquired frames, it was noticed that the `white' area fluctuates due to scattering. These aerial fluctuations are shown in Figure~5(b), with a scatter of $\approx \delta A=\pm 45~$pixels$^2$. For an average radius $r\approx 32~$pixels and an average area $A= \pi r^2$, the uncertainty in the radius is $\delta r=r\, \delta A/ (2\,A)\approx 0.2~$pixels.

In addition to the vertical motion, the oscillator also swings horizontally, making the system resemble a pendulum with periodically oscillating length, a so-called parametric oscillator \cite{case}. In Figure~\ref{fig:underdamped}(a), we depict the horizontal displacement superposed on top of the vertical oscillations. The motion for such a system deviates from simple harmonic and in fact becomes nonlinear, governed by nonlinear differential equations (e.g., Equations 1,4,6 in \cite{belyakov}) that are solved numerically. Then our predicted solution \eqref{eq:xdamp} should only be approximate. However, the excellent curve fit indicates that the nonlinearities are indeed small and the damped oscillations represent, to a fairly good extent, en exponentially damped sinusoid.

 The horizontal amplitudes are small when compared with the vertical ones, approximately one part in ten ($10\%$). We can use these small horizontal amplitudes, $\delta x$, to estimate the uncertainty in the vertical coordinates, $\delta y$, under certain simplifying assumptions. Suppose $\theta$ is the angle subtended by the horizontal swing of the pendulum from the vertical. If $l_0$ is the mean length of the spring-mass system and $l_0\pm \Delta l_{max}$ are the length extrema, the angles at the extrema are given by $\theta=\tilde{x} / (l_0\pm \Delta l_{max})$ and for small $\theta$, we have $\tilde{x}\approx x$, $\theta\approx x/(l_0\pm \Delta l_{max})$. When both the spring extension and the horizontal swings are at their maximums, the centroid-determination algorithm records a vertical position $y=(l_0+\Delta l_{max})\cos{\theta}$ whereas the position in the absence of horizontal motion would be $l_0+\Delta l_{max}$, the deviation being $\delta y=(l_0+\Delta l_{max})(\cos{\theta}-1)=(l_0+\Delta l_{max})(2\sin^2{\theta}/2)\approx(l_0+\Delta l)({\theta^2}/2)$. With the small angle approximation,
 \begin{equation}\label{eq:erroriny}
\delta y \approx \frac{(\delta x)^2}{l_0+ \Delta l_{max}},
 \end{equation}
 and similarly, for the maximum compression,
  \begin{equation}\label{eq:erroriny2}
\delta y \approx \frac{(\delta x)^2}{l_0- \Delta l_{max}}.
 \end{equation}
With ball-park figures of $\l_0\approx 200$, $\Delta l_{max}\approx 20$ and $\delta x\approx 2~$ (all units in pixels), the uncertainty $\delta y$ is $< 0.02~$pixels. In practice, the horizontal and vertical displacements are not in phase and the actual uncertainties will lie in the range specified by \eqref{eq:erroriny} and \eqref{eq:erroriny2}. Therefore, based on the light scattering and horizontal motion, a conservative guess on the order of magnitude of the positional uncertainty would be $\pm 1$ pixels.
%

\section{\label{sec:conclusions}CONCLUSIONS}

In this experiment, we have used a new technique of employing a webcam to acquire time stamped frames of a mass oscillating inside a viscous fluid. Our students take keen interest in employing the webcam, generally a household object of entertainment value, for a scientific pursuit. The experiments are low cost, ideal for cost-conscious experimental labs and easily reproducible. Furthermore, the experimentally determined viscosities of water, ethanol and methanol are in excellent agreement with previously reported values. Notably, our method returns a value far more accurate than a similar experiment that
employed a force sensor \cite{alexander}.

The experiment introduces students to damping and simple harmonic motion by acquiring data remotely. Students learn to work in pixel co-ordinates, perform simple image processing and computational tasks and test experimental observations with quantitative predictions, in our case, solutions of second order differential equations. One can also use our webcam-based approach to study the phenomenon of parametric resonance \cite{cayton}. Thus we can oscillate the combined mass-spring and pendulum system with different excitation frequencies and study the (in)stability properties of the nonlinear system as well as its transition into chaos. This is an interesting idea to pursue for short-term research or a lab project.

The experiment has been performed by a freshman class of about 150 students and very well received. It is hoped that the new method of analyzing simple harmonic motion will be welcomed by physics teachers. The authors like to thank Waqas Mahmood, Muhammad Wasif and Umer Suleman for demonstrating the experiment to the class.


\begin{thebibliography}{5}
\bibitem{nakroshis} P. Nakroshis, M. Amoroso, J. Legere and C. Smith, Am. J. Phys. \textbf{71}, 568-573 (2003).
    \bibitem{jia} D. Jia, J. Hamilton, L. M. Zaman and A. Goonewardene, Am. J. Phys. \textbf{75}, 111-115 (2007).
    \bibitem{newburgh} R. Newburgh, J. Peidle and W. Rueckner, Am. J. Phys. \textbf{74}, 478-481 (2006).
\bibitem{gilson} J. E. Gilson and O. A. Boedtker, Am. J. Phys. \textbf{37}, 1157-1158 (1969).
\bibitem{mcinerney} M. F. McInerney, Am. J. Phys. \textbf{53}, 991-996 (1985).
\bibitem{allen} M. Allen and E. J. Saxl, Am. J. Phys. \textbf{40}, 942-944 (1972).
\bibitem{gupta} V. K. Gupta, G. Shanker and N. K. Sharma, Am. J. Phys. \textbf{54}, 619-622 (1986).
\bibitem{alexander} P. Alexander and E. Indelicato, Eur J. Phys. \textbf{26}, 1-10 (2005).
\bibitem{chung} P. L. Chung, Rev. Sci. Instrum. \textbf{44}, 1669-1670 (1973).
\bibitem{pecori} B. Pecori, G. Torzo and A. Sconza, Am. J. Phys. \textbf{67}, 228-235 (1999).
\bibitem{nedev} S. Nedev and V. C. Ivanova, Eur. J. Phys. \textbf{27}, 1213-1219 (2006).
\bibitem{gil} S. Gil, H.~D. Reisin and E.~E. Rodriguez, Am. J. Phys. \textbf{74}, 768-775 (2006).
\bibitem{french} A. P. French, \textit{Vibrations and Waves} (W. W. Norton $\&$ Company, New York, 1971), 3rd ed.
\bibitem{lifshitz} L. D. Landau and E. M. Lifshitz, \textit{Fluid Mechanics}, 2nd ed. (Pergamon,
New York, 1987).
\bibitem{leshansky} A. M. Leshansky and J. F. Brady, Phys. Fluids \textbf{16}, 843-844 (2004).
\bibitem{shantarng} S. T. Chen and S. C. Wang, Physica Scripta \textbf{70}, 349-353 (2004).
\bibitem{matlab} See, for example, $\langle mathworks.com/applications/imageprocessing\rangle$.
\bibitem{main} I. G. Main, \textit{Vibrations and Waves in Physics} (Cambridge U.P., Cambridge, 1992), 3rd ed.
\bibitem{landolt} C. Wohlfarth, \textit{Landolt-Börnstein - Group IV Physical Chemistry
Numerical Data and Functional Relationships in Science and Technology} (Springer, Berlin-Heidelberg, 2008).
\bibitem{xiang} H. W. Xiang, A. Laesecke and M. L. Huber, J. Phys. Chem. Ref. Data \textbf{35}, 1597-1620 (2006).
\bibitem{case} W.~B. Case, Am. J. Phys. \textbf{64}, 215-220 (1996).
\bibitem{belyakov} A.~O. Belyakov, A.~P. Seyranian and A. Luongo, Physica D \textbf{238}, 1589-1597 (2009).
\bibitem{cayton} T.~E. Cayton, Am. J. Phys. \textbf{45}, 723-732 (1977).
\end{thebibliography}
\end{document}